\documentclass{JHEP3}

\usepackage{subfigure}
\usepackage{epsfig}
\usepackage{graphicx}
\usepackage{enumerate}
\usepackage{amsmath}
\title{A black hole instability in five dimensions?}

\author{Donald Marolf  \\
Physics Department, UCSB, Santa Barbara, CA 93106
and \\
Perimeter Institute, 31 Caroline ST. N, Waterloo, Ontario N2L 2Y5
\texttt{marolf@physics.ucsb.edu}}

\author{Amitabh Virmani \\
Physics Department, UCSB, Santa Barbara, CA 93106.\\
\texttt{virmani@physics.ucsb.edu}}

\abstract{We study the moduli-space scattering of a two-charge
supertube in the background of a  rotating BPS D1-D5-P  black hole
in 4+1 dimensions, extending the static analysis of Bena and Kraus
(hep-th/0402144). While the magnetic forces associated with this
motion change the details considerably, the final conclusion is
similar to that of the static analysis: we find that one can bring
the supertube to the horizon, so that the BMPV black hole and the
supertube merge. However, our analysis shows that this can occur
even at significantly larger values of the angular momentum than was
indicated by the static analysis. For a range of parameters,
conservation laws and the area theorem forbid the result of the
merger from being any single known object: neither near-extremal
black holes nor non-supersymmetric black rings are allowed. Such
results suggest that the merger triggers an instability of the
rotating D1-D5-P black hole, perhaps leading to bifurcation into a
pair of black objects.}

\date{May 2005}


\begin{document}


\section{Introduction}

One of the intriguing features of gravitational phenomena is the way
that black holes in 3+1 dimensions differ from their higher
dimensional cousins.   In 3+1 dimensions, black holes are fairly
simple and are associated with a variety of uniqueness and stability
results. For example, one important uniqueness theorem
\cite{robinson} states that all stationary vacuum spacetimes are
axisymmetric and belong to the Kerr family of solutions,
parameterized by their mass $M$ and angular momentum $J$. If we
require that these solutions be black holes, then the mass and
angular momentum are restricted by the relation $M^2 \ge |J|$.
Important stability results include various proofs of perturbative
stability (see, e.g., \cite{Chandrasekhar:1985kt} and references
therein) and the proof that black holes cannot bifurcate without
violating a form of cosmic censorship \cite{hawking}.  For Schwarzschild black holes, one may also
use the 2nd law of thermodynamics (and the classification of 3+1
black objects) to argue against such bifurcation since the resulting
black hole fragments would necessarily have smaller total horizon
area.  Such an argument might plausibly forbid certain quantum
mechanical process in addition to classical processes inherent in
the Einstein-Hilbert dynamics.

However, the story becomes more interesting in higher dimensions.
The spectrum of solutions is far richer and has proven to be full of
surprises. In $d \ge 4+1$ spacetime dimensions, gravity can produce
not only black holes, but also extended black objects like black
strings, black branes (see, e.g. \cite{Horowitz:1991cd}) and black
rings \cite{Emparan:2001wn}.  In this work, we restrict use of the
term ``hole'' to refer to objects with spherical horizons, while 4+1
black rings have horizons with topology $S^1 \times S^2$. In
addition, at least in certain cases such objects are known to be
unstable. The classic example of an instability was discovered by
Gregory and Laflamme \cite{Gregory:1993vy, Gregory:1994bj}, who
showed that black strings and branes can be dynamically unstable to
a breaking of translational symmetry along the extended directions.
There is also an indication \cite{Emparan:2003sy} that rapidly
rotating Myers-Perry black holes \cite{Myers:1986un} in $d \ge 6$
spacetime dimensions become unstable, imposing a dynamically
generated ``Kerr bound''
 analogous to the condition $M^2 \ge |J|$ in
3+1 dimensions. Emparan and Myers \cite{Emparan:2003sy} suggested
that this instability may catalyze a transition of these black holes
to a black ring or a bifurcation into a binary pair of Schwarzschild
black holes (which then decays via gravitational wave emission into
a single black hole of smaller angular momentum). Investigations
\cite{HM,SG,Morse,TW,TW2,C,ES,KS,TW3,KRev,GLP,DMnew} of these
instabilities have given us important insights into the nature of
gravity in higher dimensions. However, the general theory of
instabilities remains to be fully understood.

The complicated nature of some higher dimensional solutions can make
them difficult to analyze in detail.  As a result, it is useful to
draw inspiration from a variety of 3+1 dimensional studies, even if
they do not lead immediately to rigorous results. A classic such
analysis of Kerr solutions by Wald \cite{wald} probes aspects of
uniqueness and stability as well as cosmic censorship. Wald
considered extreme Kerr spacetimes (with $J=M^2$) to determine if
one can violate the $|J| \le M^2$ bound by, for example, dropping
spinning test bodies with large spin to mass ratio into such holes
(see \cite{Veronika} for an interesting discussion involving
non-extremal holes and using charges instead of angular momentum).
Note that, in the limit where an axisymmetric set of test bodies is
used, it is clear that radiation of angular momentum cannot be
relevant to this process.

Wald's results showed that no such violation occurs because the
gravitational spin-spin repulsive force prevents such a spinning
object from entering the hole.   Had the opposite result been
obtained, one would have been forced to conclude that throwing
such an object at the black hole catalyzes a transition from a Kerr black hole to
something else:  either a naked singularity (violating cosmic
censorship) or some new black hole solution (which does not exist).

Bena and Kraus \cite{Bena:2004wt} recently considered a similar
gedanken experiment in 4+1 dimensions. Their analysis used a
two-charge supertube \cite{Mateos:2001qs, Emparan:2001ux,
Mateos:2001pi} as a probe of the five dimensional rotating BPS black
hole \cite{Tseytlin:1996as, Breckenridge:1996is} (i.e., saturating a
BogomolÕnyi-Prasad-Sommerfield (BPS) bound). This black hole is
often referred to as the BMPV black hole after the authors of
\cite{Breckenridge:1996is}. The solution has three independent
charges and self-dual angular momentum, so that the magnitude of the
angular momenta in the two independent planes must be equal $(|J_1|
= |J_2|=J)$. The magnitude of the angular momentum is also bounded
by a function of the charges: $J^2 \le N_{D1} N_{D5} N_{p}$, where
$N_{D1},N_{D5},N_{p}$ are quantized charges normalized to take
integer values. While this constraint on angular momenta is relaxed
for non-extreme black holes \cite{CY1,CY2},  the difference $|J_1| -
|J_2|$ is nevertheless bounded by a function that may be taken to
parameterize the departure from extremality.

Bena and Kraus \cite{Bena:2004wt} considered static supertube
configurations in the BMPV background and found that some such
configurations exist arbitrarily close to the black hole horizon.
This suggested that one can dynamically merge the supertube and
black hole into a single object. Since supertubes carry unequal angular momentum in the
two planes, they argued that the result of the collision should be
some new  supersymmetric black hole with $|J_1| \neq |J_2|$.  In
contrast, they also argued that supertubes with large angular
momentum (relative to the scale set by the background) are prevented
from reaching the horizon, so that one is protected against forming
a final object which violates the constraint $J^2 \le N_{D1} N_{D5}
N_{p}$.

In this work we again consider the collision of a two-charge
supertube with a BMPV black hole. We extend the analysis of Bena and
Kraus beyond the static case to address slowly moving supertubes in
a BMPV background using the moduli space approximation.  Thus, we
include the effects of various velocity-dependent forces (i.e.,
magnetic forces) as well as the small departures from the BPS limit
inherent in any dynamic process.  Our methods are similar to those
of \cite{PZ}, who consider the moduli-space scattering of objects
dual to supertubes in a field theoretic (non-gravitating)
description.

The details of the dynamics are quite different from those which
might be inferred from the static analysis.  We find that magnetic
forces cause {\it any} BPS supertube configuration (having the same
symmetries assumed in \cite{Bena:2004wt}) to be separated from the
horizon by a wall in an effective potential, so that giving this
supertube any sufficiently small velocity fails to result in a
merger with the black hole.  However, the potential always
approaches its BPS value at the horizon.  Thus, additional classical
forces can lift the supertube over the wall (or it may quantum
mechanically tunnel through the wall), leading to merger with the
black hole with any arbitrarily small energy above the BPS bound.
The end result is therefore that, in agreement with Bena and Kraus
\cite{Bena:2004wt}, one can indeed bring the supertube to the
horizon of the black hole so that the two objects merge. However, in
contrast to \cite{Bena:2004wt} we find that it  is {\it also}
possible to violate even the non-extreme version of the constraint
$J^2 \le N_{D1} N_{D5} N_{p}$.

Due to the kinetic energy inherent in our scattering, our final
object will not saturate a BPS bound.   However, one can tune
parameters to make the amount of non-extremality arbitrarily small.
With such tuning, and in certain parameter regimes, conservation
laws and the area theorem prevent the final object from being either
a known non-extreme black hole \cite{CY1,CY2} or any small
deformation of the known BPS black rings \cite{Ring1, Ring2, Ring3}
(e.g., a non-BPS black ring \cite{NSring}).

However, the literature does now contain BPS solutions with the
correct conserved quantities and having a larger area than the
original BMPV black hole.  These solutions are the BPS {\it pairs}
of black rings  studied in \cite{Gauntlett:2004wh,
Gauntlett:2004qy}. Thus, some non-extreme deformation of these
solutions could represent the final result of our collision.  Here
we note the sharp contrast with the 3+1 dimensional case in which
splitting a black hole into fragments always reduces the total
entropy; in the context of \cite{Gauntlett:2004wh, Gauntlett:2004qy}
the break-up of BMPV into two or more black rings is entropically
favored, and our collision could well trigger an instability of BMPV
leading to such a fragmentation. The picture is similar to one of
the scenarios discussed by Emparan and Myers in
\cite{Emparan:2003sy}, except that our process would occur in 4+1
dimensions and we have studied the regime close to the BPS limit (as
opposed to the uncharged regime). Note that the solutions of
Gauntlett and Gutowski \cite{Gauntlett:2004wh, Gauntlett:2004qy}
were unknown at the time of \cite{Bena:2004wt}, and even the
existence of supersymmetric black rings was in question.

Our discussion below is organized as follows. We begin with a review
in section \ref{prelims} of relevant properties of BMPV black holes
and some other black objects. Then, in section \ref{mod}, we set up
the formalism for a supertube moving in the (type IIA dual of the) BMPV background and
compute the associated action in the moduli space approximation.
Some details are relegated to appendix \ref{appendix}. In section
\ref{scatter} we study the detailed dynamics of the supertube and
find in our approximation that a supertube with arbitrary charges
can merge with the BMPV black hole. We  conclude in section
\ref{discussion} with some discussion and further consideration of the possible end results of our collision.


\section{BMPV black holes and other black objects}\label{prelims}

In this section we review certain properties of BMPV black holes
\cite{Tseytlin:1996as, Breckenridge:1996is} (see also
\cite{Herdeiro:2000ap} for a useful representation) and other
relevant black objects. A BMPV black hole  is a supersymmetric,
rotating, asymptotically flat solution of the $T^5$ reduction of
Type IIB supergravity to five dimensions. The lift of the BMPV
solution to ten dimensions is given in the string frame by
\begin{eqnarray}
\nonumber ds^2 &=& H_{D1}^{-1/2} H_{D5}^{-1/2} \bigg{[}-dt^2 + dz^2
+ (H_{p} -1)(dt - dz)^2  + 2 (\gamma_1 \theta d \phi_1 + \gamma_2
\theta d \phi_2)(dz - dt) \bigg{]}
\\[2mm]
\label{bmpv-field1}
&+&H_{D1}^{1/2}H_{D5}^{1/2}(ds^2_{\mathbb{R}^4}) +
H_{D1}^{1/2}H_{D5}^{-1/2}(ds^2_{{T}^4}),
\\[2mm]
\label{bmpv-field2} C_\mathit{2} &=&  (H_{D1}^{-1}-1) dt \wedge dz -
r_{D5}^2 \cos^2 \theta d\phi_1 \wedge d \phi_2 + H_{D1}^{-1}(dt
-dz)\wedge (\gamma_1 d\phi_1 + \gamma_2
d\phi_2) \\[2mm]
\label{bmpv-field3} e^{2 \Phi} &=& \frac{H_{D1}}{H_{D5}},
\end{eqnarray}
where
\begin{equation}
\gamma_1 = \frac{\omega}{r^2} \sin^2 \theta \quad \mbox{ and } \quad
\gamma_2 = \frac{\omega}{r^2} \cos^2\theta,
\end{equation}
and where the fields are normalized as in \cite{Joe}. The solution
is specified by flat metrics ($ds^2_{\mathbb{R}^4}$ and
$ds^2_{{T}^4}$) on $\mathbb{R}^4$ and $T^4$, three Harmonic
functions ($H_{D1}, H_{D5}$ and $H_{p}$) and \emph{only one} angular
momentum parameter $\omega$.  Although the solution has angular
momenta $(J_1,J_2)$ in two independent planes (associated with the
commuting Killing vectors $\partial_{\phi_1},
\partial_{\phi_2}$), the solution satisfies $J_1= J_2=\frac{\pi}{4G_5} \omega$.  The
internal Euclidean space, $T^4$, is parameterized by Cartesian
coordinates $x_6, x_7, x_8$ and $x_9$ while the external Euclidean
space, $\mathbb{R}^4$, has coordinates $r,\theta, \phi_1, \phi_2$
with metric
\begin{equation}\label{metric}
ds^2_{\mathbb{R}^4} = dr^2 + r^2 d\theta^2 + r^2 \sin^2 \theta
d\phi_1^2 + r^2 \cos^2 \theta d\phi_2^2.
\end{equation}
Here $r$ is the radial coordinate and $\theta, \phi_1$ and $\phi_2$
are angles on $S^3$. These coordinates ($r, \theta, \phi_1, \phi_2$)
can be obtained from Cartesian coordinates $x_1,x_2,x_3,x_4$ on
$\mathbb{R}^4$ through:
$$
x_1 + i x_2  = r \sin \theta e^{i \phi_1}; \quad  x_3 + i x_4 = r
\cos \theta e^{i \phi_2},
$$
where $\theta $ ranges from $0$ to $\pi/2$ and $\phi_1, \phi_2$
ranges over $[0, 2\pi)$. The three Harmonic functions ($H_{D1},
H_{D5}$ and $H_{p}$) on the external Euclidean space
$\mathbb{R}^4$ are given by
\begin{equation} H_{D1} = 1 +
\frac{r_{D1}^2}{r^2}; \quad H_{D5} = 1 +
\frac{r_{D5}^2}{r^2}; \quad H_{p} = 1 +
\frac{r_{p}^2}{r^2},
\end{equation}
and the coordinate singularity at $r=0$ is a smooth horizon of the
black hole.

Compactifying the $z$-direction on $S^1$ yields a five dimensional
black hole solution with three distinct charges. Denoting the
asymptotic length of the $S^1$ by $2\pi R_{B}$ and the volume of the
$T^4$ by $(2\pi \ell)^4 $, the quantized integral charges are
\cite{malda} :
\begin{equation}\label{quant1}
N_{1} = \frac{\ell^4}{g_{B} \alpha'^3} r_{D1}^2,  \qquad N_{5} =
\frac{r_{D5}^2}{g_{B}\alpha'},  \qquad N_{P} = \frac{R_B^2 \ell^4
}{g_{B}^2 \alpha'^4}r_{p}^2.
\end{equation}
where $g_B$ is the string coupling. Here the $B$ subscripts refer to
the fact that (\ref{bmpv-field1}) -- (\ref{bmpv-field3}) is a type
IIB solution and foreshadow the fact that we will shortly dualize
this solution to a IIA frame.

The ADM mass, angular momenta and the entropy of the corresponding
five dimensional solution are
\begin{eqnarray}
M_{BMPV} &=& \frac{\pi}{4 G_5}\left( r_{D1}^2 + r_{D5}^2 + r_{p}^2 \right) = \frac{1}{g_{B}^2} \left( \frac{R_B g_{B} N_{D1}}{\alpha'} + \frac{R_B \ell^4 g_{B} N_{D5}}{\alpha'^3} + \frac{g_B^2 N_{p}}{R_B}\right),\\
J_{\phi_1} &\equiv& J_1 = J = \frac{\pi}{4 G_5} \omega, \quad J_{\phi_2} \equiv J_2 = J = \frac{\pi}{4 G_5} \omega,  \\
S_{BMPV} &=& \frac{2\pi^2}{4 G_5} \sqrt{r_{D1}^2 r_{D5}^2 r_{p}^2 -
\omega^2} = 2 \pi \sqrt{N_{D1} N_{D5} N_{p} - J^2}.
\end{eqnarray}
Here the five dimensional Newton's constant $G_5$ is related to ten
dimensional Newton's constant $G_{10} = 8 \pi^6 \alpha'^4 g_{B}^2$
by
$$
G_5 = \frac{G_{10}}{2 \pi R (2 \pi \ell)^4} = \frac{\pi \alpha'^4
g_B^2}{4 \ell^4 R_B}.
$$
The solution obeys a Kerr like bound
\begin{equation}\label{kerr-like}
\omega^2 \le r_{D1}^2 r_{D5}^2 r_{p}^2 \mbox{\ or, \  equivalently,
 }J^2 \le N_{D1} N_{D5} N_{p}.
\end{equation}
Note that we have set $\hbar=1$ so that the angular momentum $J$
takes half-integer values.  When the bound (\ref{kerr-like}) is
violated, the horizon disappears to expose naked closed time-like
curves.


The solution (\ref{bmpv-field1})--(\ref{bmpv-field3}) will be of
central use in section \ref{mod} below, where we study the moduli
space scattering of supertubes in this background.  However, we will
also be interested in a few basic features of other solutions, such
as the non-BPS version of (\ref{bmpv-field1})--(\ref{bmpv-field3})
found by Cvetic and Youm \cite{CY1,CY2}. These non-BPS solutions are
parameterized by three charges and also by three additional
parameters $m$, $L_1$, and $L_2$ which are related respectively to
the energy above extremality and to the two independent angular
momenta.  We will be most interested in the near-BPS limit, also
studied in \cite{CY1,CY2}. In this limit $m \ll
r^2_{D1},r^2_{D5},r^2_{p}$ and the ADM mass $M_{CY}$ for the
Cvetic-Youm solutions is given \cite{CY1,CY2} to leading non-trivial
order in $m$ by
\begin{eqnarray}
M_{CY} & \approx & M_{BMPV} + \frac{\pi}{4 G_5} \frac{m^2}{2} \left(
\frac{1}{r_{D1}^2} + \frac{1}{r_{D5}^2} + \frac{1}{r_{p}^2}\right) +
\mathcal{O}(m^2).
\end{eqnarray}
The (half-integer) angular momenta are determined by the
relations\footnote{Note that we have chosen a convention so that
$J_1$ and $J_2$ have no relative minus sign in the BPS limit. This
differs from the conventions of \cite{CY1, CY2}.}
\begin{equation}
\label{J+} \frac{J_1 + J_2}{2} = \frac{\pi}{4 G_5 \sqrt{2}} r_{D1}
r_{D5} r_{p} (L_1 + L_2) + \mathcal{O}(m^2)
\end{equation}
and
\begin{equation}
\label{J-} \frac{J_1 - J_2}{2} = \frac{\pi}{4 G_5 \sqrt{2}} r_{D1}
r_{D5} r_{p} \left( \frac{1}{{r}_{D1}^2} + \frac{1}{r_{D5}^2} +
\frac{1}{r_{p}^2}\right)(L_1 - L_2) + \mathcal{O}(m^2),
\end{equation}
while the integer charges $N_{D1}, N_{D5}, N_p$ are again given by
(\ref{quant1}) and, to leading non-trivial order in $m$, the entropy
is given by
\begin{eqnarray}
\label{2.13} S_{CY} &\approx& S_{BMPV} + \frac{m  \pi^2}{4 G_5}
\frac{ r_{D1}^2  r_{D5}^2 + r_{D5}^2 r_{p}^2 + r_{D1}^2 r_{p}^2
}{r_{D1} r_{D5} r_{p}} \sqrt{1- \frac{1}{2}(L_1 -L_2)^2} +
\mathcal{O}(m^2),
\end{eqnarray}
where we have corrected a minor error in the expansion given in \cite{CY1, CY2}.
It is important to note that, although these solutions carry two
independent angular momenta in orthogonal planes, their difference is
bounded and satisfies
\begin{equation}\label{bound}
|J_1 - J_2| \le \frac{\pi m}{ 2 G_5} r_{D1} r_{D5} r_{p} \left(
\frac{1}{r_{D1}^2} + \frac{1}{r_{D5}^2} + \frac{1}{r_{p}^2}\right) +
\mathcal{O}(m^2).
\end{equation}


In addition to the above black holes, a new family of supersymmetric
black solutions with the same asymptotic charges and horizon
topology $S^1\times S^2$ has recently been constructed in
\cite{Ring1, Ring2, Ring3}. These ``black ring'' solutions are
labeled by 7 independent parameters. They carry three charges,
$N_{D1}$, $N_{D5}$ and $N_{p}$ (related to 3 charge radii $r_{D1}$,
$r_{D5}$, $r_{p}$), two independent angular momenta $J_1$ and $J_2$
and three dipole charges $n_{D1}, n_{D5}$ and $n_{KK}$.  The dipole
charges are not conserved. The ADM mass and angular momenta take the
following form
\begin{eqnarray}
\label{ringrels}
M_{ring} &=& \frac{\pi}{4 G_5}\left( r_{D1}^2 + r_{D5}^2 + r_{p}^2 \right) =
\frac{1}{g_{B}^2} \left( \frac{R_B  g_{B} N_{D1}}{\alpha'} + \frac{R_B \ell^4 g_{B} N_{D5}}{\alpha'^3}
+ \frac{g_B^2 N_{p}}{R_B}\right),\\
J_2 &=& \frac{1}{2} \left( n_{D1} N_{D5} + n_{D5} N_{D1} + n_{KK}
N_{P} - n_{D1} n_{D5}
n_{KK} \right), \\
J_1 &=& J_2 + \frac{\pi R^2}{4 G_5} \left(\frac{g_B \alpha'^3}{R_B
V}n_{D1} + \frac{g_B \alpha'}{R_B} n_{D5}  +  R_B n_{KK} \right),
\end{eqnarray}
where $R$ is a parameter appearing explicitly in the solution and is
called the ``radius'' of the ring.   Note that, as we will later be
interested in rings for which the larger angular momentum lies in
the $\phi_1$ plane, we have permuted the angular momenta relative to
the conventions of  \cite{Ring1, Ring2, Ring3}.  The entropy of the
black ring is given by
\begin{equation}
\label{Sring} S_{ring} = 2 \pi \sqrt{N_{D1} N_{D5} N_{p} - \mathcal{N}_{D1}
\mathcal{N}_{D5} \mathcal{N}_{p} - J_2^2 - n_{D1} n_{D5} n_{KK} (J_1 -
J_2)},
\end{equation}
where we have used the notation $\mathcal{N}_{D1} := N_{D1} - n_{D1}
n_{KK}$, $\mathcal{N}_{D5} := N_{D5} - n_{D5} n_{KK}$ and
$\mathcal{N}_{p} := N_{p} - n_{D1} n_{D5}$.  We have normalized both
the monopole charges $N_{D1},N_{D5},N_p$ and the dipole charges
$n_{D1}, n_{D5}, n_{KK}$ so that they take integer values and, as
usual, the angular momenta $J_1,J_2$ take half-integer values.

In the limit $R\rightarrow0$ the black ring solution reduces to the
BMPV solution. However, the horizon area is discontinuous at $R=0$,
where the horizon changes topology from the  $S^1 \times S^2$ of the
black ring to the $S^3$ of the BMPV black hole.  An analysis of
(\ref{Sring}) shows \cite{Ring3} that the  $R\rightarrow 0$ limit of
the black ring horizon area is always less than that of the
corresponding BMPV black hole.  One may also show by further
calculation that, for fixed asymptotic charges $N_{D1},N_{D5},N_{p}$, the
entropy of the black ring is maximized for
\begin{equation}
\label{maxSq}
n_{D1}= \sqrt{\frac{N_{D1} N_{p}}{3N_{D5}}}; \quad n_{D5} =
\sqrt{\frac{N_{D5} N_{p}}{3N_{D1}}}; \quad n_{KK}= \sqrt{\frac{N_{D1} N_{D5}}{3N_{p}}} \mbox{ and } J_1 = J_2 = \frac{4 \sqrt{N_{D1} N_{D5}
N_{p}}}{3 \sqrt{3}}.
\end{equation}
This maximal entropy is given by
\begin{equation}\label{max-entropy}
S_{ring}(max) = 2 \pi \frac{\sqrt{N_{D1} N_{D5} N_{p}}}{3}.
\end{equation}
Since the two angular momenta (\ref{maxSq}) are equal, the configuration is not
really a black ring, but the result (\ref{max-entropy}) does provide
an upper bound on the entropy of all black rings with fixed
$N_{D1},N_{D5},N_{p}$.  Note that (\ref{max-entropy}) is also the entropy of
a BMPV black hole with a larger angular momentum ($J = \frac{2
\sqrt{2}}{3} \sqrt{N_{D1} N_{D5} N_{p}}$).

A particularly interesting feature of BPS black rings is that one
can sometimes increase their entropy \cite{Gauntlett:2004wh}  by
splitting them into several BPS black rings with smaller values of
the quantized charges $N_{D1},N_{D5},N_{p}$.  In particular, the results of
Gauntlett and Gutowski \cite{Gauntlett:2004wh} indicate that for any  BMPV
black hole with $J > 0$, there are pairs of black rings whose
asymptotic charges and angular momenta sum to the same values as those of the
black hole and for which the sum of the black ring horizon areas is
{\it larger} than that of the BMPV black hole.  The upper bound on
the entropy of a single black ring mentioned above thus means that
one can also increase the entropy of black rings with small radius
$R$ by replacing them with a pair of black rings having the same
total asymptotic charges and angular momenta.  Thus, at least at
small radius $R$ one might expect black rings to be unstable to
fragmentation.

\section{Supertubes and the action on moduli space}\label{mod}

In this section we review certain properties of two-charge
supertubes (section \ref{2q}) and obtain the moduli space action
(section \ref{moduli}) for a slowly moving supertube in the BMPV
background.

\subsection{Two-charge supertubes}
\label{2q}

As with all stringy objects, supertubes admit a variety of dual
descriptions.  We may therefore choose any duality frame which is
convenient for the purpose at hand.  Our goal below is to describe
the supertube dynamics in a simple and familiar way.  We therefore
choose the IIA duality frame in which the tube carries D0 and F1
charges, as in this frame the tube admits a description in terms of
the D2-brane Born-Infeld action.  Such a description was used in the
original work \cite{Mateos:2001qs, Emparan:2001ux, Mateos:2001pi},
in which the supertubes are described as cylindrical D2 branes
supported against collapse by the angular momentum associated with
crossed electric and magnetic fields (which account for the F1 and
D0 charges).  See also \cite{NO1,NO2,NO3}.

We remind the reader that, at each time, such tubes have topology
$S^1_{cross} \times \mathbb{R}$ or, if they wrap a compact $S^1$,
they have topology $S^1_{cross} \times S^1_z$.  Here we have
distinguished the factor $S^1_{cross}$ (the ``cross-section'' of the
tube) from the factor $S^1_z$ which is always associated with a
translational symmetry.  Below, this symmetry corresponds to shifts
of the coordinate $z$.  The factor $S^1_z$ is also the direction
along which the F1 strings are extended. When propagating in
Minkowski space, the cross section of a supertube can be an
arbitrary (non-intersecting) closed curve; all such configurations
saturate a BPS bound associated with their F1 and D0 charge.  Due to
their topology, such supertubes carry no net D2-brane charge, but
they do carry a D2 dipole moment.

The Born-Infeld action for a D2 brane is given by
\begin{eqnarray}
\label{DBI} S &=& S_{DBI} + S_{WZ} = \int {\cal L} d^3x \cr
\label{action} &=& - \tau_{D2} \int d^3 x e^{-\Phi} \sqrt{-det(g_{ab} +
b_{ab} + F_{ab})} + \tau_{D2} \int c \wedge e^{F_\mathit{2} +
b_{\mathit{2}}}
\end{eqnarray}
where $g_{ab}$ and $b_{ab}$ are respectively the metric and the
Neveu-Schwarz two-form field induced on the supertube by immersion
in some background.  Similarly, $c_\mathit{1}, c_{\mathit{3}}$ etc.
are the fields induced on the supertube by the background
Ramond-Ramond fields and $\Phi$ is the background dilaton.  Finally,
$F = \frac{1}{2}F_{ab} dx^a \wedge dx^b$ is the
D2-brane world-volume gauge-field and $\tau_{D2}$ is the D2-brane
tension $\tau_{D2} = \frac{1}{(2 \pi)^2 g_{A} \alpha'^{3/2}}$, where
$g_A$ is the string coupling in our present type IIA duality frame.

Due to the gauge field $F_{ab}$, the D2-brane will carry D0-brane
charge and F1-string charges:
\begin{equation}\label{charges}
q_{D0} = \frac{\tau_{D2}}{\tau_{D0}} \int dz \int d\sigma F_{\sigma z}
\quad \mbox{and} \quad q_{F1} = \frac{1}{\tau_{F1}} \int d\sigma
\frac{\partial {\cal L}}{\partial F_{tz}},
\end{equation}
where $\tau_{D0} = \frac{1}{g_A \sqrt{\alpha'}}$ and $\tau_{F1}  =
\frac{1}{2 \pi \alpha'}$ are the D0 and F1 brane tensions
respectively. We have normalized $q_{F1}$ and $q_{D0}$ so that they
take integer values.

Supertubes are supersymmetric configurations, which we may take to
be time-independent.  Let us briefly review the properties of the
original supertubes \cite{Mateos:2001qs} associated with embedding
the D2-brane in Minkowski space.  Such configurations have gauge
fields of the form \cite{Mateos:2001qs}
\begin{equation}
\label{staticF} F = F_{tz} dt \wedge dz + F_{\sigma z}
d\sigma \wedge d z,
\end{equation}
where $\sigma$ is a coordinate on the $S^1$ cross-section and $z$ is
a coordinate along the $\mathbb{R}$ direction.  In particular,
supersymmetry requires $F_{tz}=1$ in string units (with $2 \pi
\alpha' =1$). However, the magnetic field  $F_{\sigma z}$ can be an
arbitrary non-vanishing function of $\sigma$.  We take $F_{\sigma
z}$ to be positive below.  Under such conditions, the energy
constructed from (\ref{action}) saturates a BPS bound:
\begin{equation}\label{bps}
E = \tau_{D0} |q_{D0}| + \tau_{F1}|q_{F1}| 2 \pi R_A,
\end{equation}
where $ 2 \pi R_A$ is the length $S^1_z$. This result shows that the
positive energies associated with the D2-brane tension and with the
rotation are exactly canceled by the negative binding energy of D0
and F1 branes and suggests that supertubes may be interpreted as
marginal bound states\footnote{See \cite{MCP} for a demonstration in
this duality frame that the BPS states described above are in fact
bound; i.e., that the spectrum of states is discrete despite the
arbitrary shape of $S^1_{cross}$. See also \cite{NOS}.} of D0 branes
and F1 strings.  If one also imposes rotational symmetry around the
$S^1$, then a singly wound such supertube has angular momentum $j =
{q}_{F1} {q}_{D0}$ in the plane of the $S^1$.

\subsection{Moduli space action}\label{moduli}

Our goal is to study the moduli-space motion of a supertube in the
BMPV black hole background. However, we have presented the supertube
above in a IIA duality frame while the BMPV background
(\ref{bmpv-field1}--\ref{bmpv-field3}) was given in a IIB frame.  As
a result, it is convenient to T-dualize the BMPV background
(\ref{bmpv-field1})--(\ref{bmpv-field3}) along the D1-branes
following \cite{Buscher:1987sk,Buscher:1987qj, RRTdual}. The result
is:
\begin{eqnarray}\label{t-dual-metric}
\nonumber ds^2 &=& - H_{D0}^{-1/2}H_{D4}^{-1/2} H_{F1}^{-1} \left(dt
+ \gamma_1 d\phi_1 + \gamma_2
d\phi_2 \right)^2 +  H_{D0}^{1/2}H_{D4}^{1/2} H_{F1}^{-1} dz^2
\\[2mm]
&+& H_{D0}^{1/2}H_{D4}^{1/2} (dr^2 + r^2 d\theta^2 + r^2 \sin^2
\theta d\phi_1^2 + \cos^2 \theta d\phi_2^2) +
H_{D0}^{1/2}H_{D4}^{-1/2}ds^2_{T_4}
\end{eqnarray}
while the dilaton and the gauge fields are
\begin{eqnarray}\label{dilaton}
e^{2 \Phi} &=& \frac{H_{D0}^{3/2}}{H_{D4}^{1/2} H_{F1}} \\[2mm]
\label{c1} C_\mathit{1} &=& (H_{D0}^{-1} -1) dt  + H_{D0}^{-1}
(\gamma_1 d \phi_1 + \gamma_2 d\phi_2)
\\[2mm] \label{c3}
C_\mathit{3} &=& -(H_{D4} -1)r^2 \cos^2 \theta d\phi_1 \wedge d
\phi_2 \wedge d z + H_{F1}^{-1} dt\wedge (\gamma_1 d\phi_1 +
\gamma_2 d\phi_2) \wedge dz
\\[2mm] \label{b2}
B_{\mathit{2}} &=& (H_{F1}^{-1} -1) dt \wedge dz + H_{F1}^{-1}
(\gamma_1 d\phi_1 + \gamma_2 d\phi_2) \wedge dz.
\end{eqnarray}
In (\ref{t-dual-metric})--(\ref{b2}),
 we have introduced $H_{D0} = H_{D1}, H_{D4} = H_{5}, H_{F1} =
H_{p}$ with $r_{D0} = r_{D1}, r_{D4} = r_{D5}, r_{F1} = r_{p}$ in
order to reflect the charge structure in the IIA frame. The
quantized charges in this duality frame are obtained by applying
T-duality to the expressions (\ref{quant1}):
\begin{equation}\label{quant2}
N_{D0} = N_{1} = \frac{\ell^4 R_A}{g_{A} \alpha'^{7/2}} {r}_{D0}^2,
\quad N_{D4} = N_{D5} = \frac{ R_A}{g_{A}\alpha'^{3/2}}{r}_{D4}^2,
\quad N_{F1} = N_{p} = \frac{\ell^4 }{g_{A}^2 \alpha'^3}r_{F1}^2.
\end{equation}
where $g_A = g_B \frac{\sqrt{\alpha'}}{R_B}$ is the Type IIA string
coupling and $R_A = \frac{\alpha'}{R_B}$.

Let us now consider test D2-branes in the background
(\ref{t-dual-metric}--\ref{b2}).  So long as the $z$ directions on
the world-volume and in the background coincide, one finds \cite{Bena:2004wt}  that
static supersymmetric tubes have properties similar
to the those in Minkowski space described in
section \ref{2q}.  In particular, if the worldvolume gauge fields again
take the form (\ref{staticF}), with $F_{tz}=1$, the configuration is
supersymmetric for any $F_{\sigma z}$ and any embedding of the $S^1$
into the external spatial directions $r, \theta, \phi_1, \phi_2$ at
a fixed value of the internal torus coordinates ($x_6,x_7,x_8,x_9$).
The energy of such tubes is again given by the BPS bound
(\ref{bps}).

Now, we wish to use a supertube to dynamically probe the IIA BMPV
background (\ref{t-dual-metric})--(\ref{b2}).  The motion of such
tubes will break supersymmetry, so that we will be forced to
generalize the static ansatz above.  However, we may simplify the
analysis by imposing translational symmetry in both $\sigma$ and
$z$. Thus, the world-volume gauge field takes the form
\begin{equation}\label{an1}
F = F_{tz} dt \wedge dz + F_{\sigma z} d\sigma \wedge d
z + F_{t \sigma} dt \wedge d \sigma,
\end{equation}
where $F_{tz}$, $F_{t\sigma}, $ and $F_{\sigma z}$ are functions of
time alone. We will work in static gauge; i.e. we take $t$ and
$z$ to coincide on the worldvolume and in the spacetime. The
supertube is then described by the ansatz
\begin{eqnarray}\label{an2}
t_{\rm spacetime} &=& t_{\rm supertube} \cr z_{\rm spacetime} &=&
z_{\rm supertube}.
\end{eqnarray}
We will also take the supertube to remain at an arbitrary (fixed) location on the $T^4$.  The rest
of the coordinates $(r, \theta, \phi_1, \phi_2)$ are functions of
$\sigma$ and $t$ alone.

The action governing the dynamics of the slowly moving supertube is
then obtained by inserting the ansatz (\ref{an1}$-$\ref{an2}) and
the IIA BMPV background (\ref{t-dual-metric}--\ref{b2}) into the
D2-brane action (\ref{action}). The dynamical variables are the
coordinates of the supertube and the deviations of the components of
the 2-form field $F_{\mathit 2}$ from their supersymmetric values.
In particular, we denote the deviation of the electric field
$F_{tz}$ by $\delta F_{tz}:= F_{tz} -1 $.

We now carry out an expansion to second order in the velocities and in the
quantities $\delta F_{tz}$ and $F_{t\sigma}$. For
the Dirac-Born-Infeld part of the action (the first term on the last
line of (\ref{DBI})), the calculation is somewhat involved and is
outlined in Appendix \ref{appendix}.  However, the remaining
(Wess-Zumino) term is more straightforward and yields:
\begin{eqnarray}\label{wess-zumino}
S_{WZ} &=& \tau_{D2} \int c_{\mathit{3}} + \tau_{D2} \int
c_{\mathit{1}}\wedge(F+b)_{\mathit{2}} \cr &=& \tau_{D2} \int
c_{\mathit{3} t \sigma z} + \tau_{D2} \int c_{\mathit{1}t}
(F+b)_{\mathit{2}\sigma z} - \tau_{D2} \int c_{\mathit{1}\sigma}
(F+b)_{\mathit{2}t z} \cr
\nonumber
&=& \tau_{D2} \int dt d \sigma
dz \bigg{[} -\delta F_{tz} H_{D0}^{-1} \gamma_\sigma + (H_{D0}^{-1}
-1 + \gamma_t H_{D0}^{-1})F_{\sigma z}  - (H_{D4} -1) r^2 \cos^2
\theta
\\[2mm]
& & \times \left( \frac{\partial \phi_1}{\partial t} \frac{\partial
\phi_2}{\partial \sigma} - \frac{\partial \phi_1}{\partial \sigma}
\frac{\partial \phi_2}{\partial t}\right) \bigg{]},
\end{eqnarray}
where
\begin{equation}\label{gamma}
\gamma_\xi := \gamma_1 \frac{\partial \phi_1}{\partial \xi} +
\gamma_2 \frac{\partial \phi_2}{\partial \xi}.
\end{equation}

Note that although (\ref{wess-zumino}) is the full Wess-Zumino part
of the action, it is linear in both the velocities and $\delta
F_{tz}$ and it is independent of $F_{t \sigma}$.  Adding the
Wess-Zumino term (\ref{wess-zumino}) to the DBI term (\ref{dbi}) we
obtain the complete action (\ref{full}) to quadratic order. The
result is:
\begin{eqnarray}\label{main}
\nonumber
S &=& \tau_{D2} \int dt dz d\sigma \Bigg{[}-  F_{\sigma z}
- H_{D4} \Delta_{\sigma t} + \Bigg{\{} \frac{H_{D4} \Delta_{\sigma
\sigma}}{F_{\sigma z}}\Bigg{\}} \delta F_{tz} + \frac{H_{D4} F_{t
\sigma}^2}{2 F_{\sigma z}} - (H_{D4} -1) r^2 \cos^2 \theta \\[2mm] & &
\times \left( \frac{\partial \phi_1}{\partial t} \frac{\partial
\phi_2}{\partial \sigma} - \frac{\partial \phi_1}{\partial \sigma}
\frac{\partial \phi_2}{\partial t}\right) + \frac{ H_{D4} H_{F1}
\Sigma }{ 2 F_{\sigma z}}\Bigg{(}\Delta_{t t} - \frac{2
\Delta_{\sigma t} \delta F_{tz}}{F_{\sigma z} } +
\frac{\Delta_{\sigma \sigma} \delta F_{tz}^2}{F_{\sigma
z}^2}\Bigg{)}
 \Bigg{]},
\end{eqnarray}
where we have introduced
\begin{eqnarray}
\Sigma &:=&  F_{\sigma z}^2 + 2 \gamma_\sigma F_{\sigma z}
H_{F1}^{-1} + H_{F1}^{-1} H_{D0} H_{D4} \Delta_{\sigma \sigma} \ \
{\rm and} \\[2mm] \label{Delta} \Delta_{\xi \eta} &:=& \frac{\partial
r}{\partial \xi}\frac{\partial r}{\partial \eta} + r^2
\frac{\partial \theta }{\partial \xi}\frac{\partial \theta}{\partial
\eta} + r^2 \sin^2 \theta \frac{\partial \phi_1}{\partial \xi}
\frac{\partial \phi_1}{\partial \eta} + r^2 \cos^2 \theta
\frac{\partial \phi_2}{\partial \xi} \frac{\partial \phi_2}{\partial
\eta}.
\end{eqnarray}

The action (\ref{main}) is the key result for the discussion below.
It is valid for an arbitrary circular embedding centered on the BMPV
black hole.   Setting velocities to zero and values of the
worldvolume fields to their supersymmetric values, the action
(\ref{main}) reduces to the result of \cite{Bena:2004wt}:
\begin{equation}
S_{\rm susy} = -\tau_{D2}\int dt dz d \sigma F_{\sigma z} = - \tau_0
\int q_{D0} dt.
\end{equation}

A quantity of great interest is the F1 charge, obtained by
differentiating the Lagrangian (\ref{main}) with respect to $F_{tz}$
(see \ref{charges}). We find
\begin{eqnarray}\label{F1-charge}
q_{F1} = \frac{\tau_{D2}}{\tau_{F1}} \int d\sigma  \Bigg{[}
\frac{H_{D4} \Delta_{\sigma \sigma}}{F_{\sigma z}} - \frac{H_{D4}
H_{F1} \Delta_{\sigma t}  \Sigma }{F_{\sigma z}^2}   +
\frac{\Delta_{\sigma \sigma} H_{D4} H_{F1} \Sigma }{F_{\sigma z}^3}
\delta F_{tz}\Bigg{]}.
\end{eqnarray}
Setting $\delta F_{tz} = 0 $ and velocities to zero leaves just the
first term in (\ref{F1-charge}), which  again reproduces the results
of \cite{Bena:2004wt}.

The dynamics of the world-volume field $F$ must ensure
conservation of  the various supertube charges.  To see this for the
D0 charge, recall that we have taken  $F$ independent of
both $\sigma$ and $z$, i.e. $\partial_z F_{ab} =
\partial_\sigma F_{ab} = 0$. The Bianchi identity
\begin{equation}
\partial_t F_{\sigma z} + \partial_z F_{t \phi} + \partial_\sigma F_{zt} = 0
\end{equation}
 then implies
\begin{equation}
 \partial_t F_{\sigma z} = 0.
\end{equation} Integrating this result over the D2-brane worldvolume yields conservation of the D0 charge.

The other conservation laws follow from the equations of motion:
\begin{equation}
\label{FEOM} 0 = \partial_a \frac{\partial {\cal L}}{\partial
F_{ab}} =  \partial_t \frac{\partial {\cal L}}{\partial F_{tb}},
\end{equation}
where in the last step we have used the assumed translation
invariance in $\sigma$ and $z$. Integrating (\ref{FEOM}) over an appropriate cross section of the D2
worldvolume yields conservation of an F1 charge in the $b$-direction
on the D2-brane.  For $b=z$, it yields $\frac{dq_{F1}}{dt}=0$, while
for $b=\sigma$ we find
\begin{equation}
\label{F1sigma} \frac{\partial {\cal L}}{\partial F_{t \sigma}} =
\tau_{D2}  \frac{H_{D4} F_{t\sigma}}{F_{\sigma z}} = \mbox{ const }.
\end{equation}
From (\ref{F1sigma}) we see that, due to our assumed translation
invariance, it is consistent to also set $F_{t \sigma}$ to zero.


\section{Scattering}\label{scatter}

In this section we use the action (\ref{main}) to understand the
dynamics of the supertube as it moves in the black hole background.
Recall that we are interested in the dependence of the result on the
angular momentum of the supertube.  In particular, we are interested
in whether the dynamics allows the supertube to merge with the black
hole when the total angular momentum does not correspond to that of
another BMPV black hole.  Note that we have assumed sufficient
symmetries to reduce the problem to five spacetime dimensions, where
angular momenta are classified by two-independent parameters
associated with the block diagonalization of the angular momentum
two-form.  It is simplest to investigate this question in the case
where the angular momenta of the BMPV black hole and the supertube
are simultaneously block diagonalizable.  Since the angular
momentum of a static circular supertube always lies in some plane, this
occurs only when the plane of the supertube coincides with one of
the principal angular momentum planes  of the black hole.  Since the
black hole is symmetric under exchange of these planes, we may as
well follow \cite{Bena:2004wt} and study embeddings with $r, \theta,
\phi_2$ independent of $\sigma$.  We fix the remaining
reparametrization freedom by requiring
\begin{equation}\label{ansatz}
\phi_1(\sigma,t) = \sigma.
\end{equation}

Such embeddings provide the $\sigma$-translation symmetry assumed
above, which in turn guarantees that this ansatz is consistent.  We
also impose $F_{t \sigma}=0$, which is consistent by
(\ref{F1sigma}).  Thus the dynamics for our ansatz follows from the
action (\ref{main}) restricted to such embeddings. The result is
(where dot denotes the derivatives with respect to time)
\begin{eqnarray}\label{embed-action}
\nonumber
S &=&  (2\pi)^2 R_A  \tau_{D2} \int dt \Bigg{[} -F_{\sigma
z} + r_{D4}^2 \cos^2 \theta \dot{\phi_2} +
\delta F_{tz} \frac{H_{D4} r^2 \sin^2\theta}{F_{\sigma z}} \\
\nonumber &+& \Bigg{\{} \frac{F_{\sigma z} H_{D4} H_{F1}}{2} +
H_{D4} \gamma_1 + \frac{H_{D0} H_{D4}^2 r^2 \sin^2 \theta}{2
F_{\sigma z}} \Bigg{\}} \\[2mm] &\times& \left( \frac{\delta F_{tz}^2 r^2
\sin^2 \theta}{F_{\sigma z}^2} + \dot{r}^2 + r^2 \dot{\theta}^2 +
r^2 \cos^2 \theta \dot{\phi_2}^2 \right) \Bigg{]}.
\end{eqnarray}
Similarly, the F1 charge of our configuration is
\begin{equation}\label{cF1-charge}
q_{F1} = 2 \pi \frac{\tau_{D2}}{\tau_{F1}}  \frac{H_{D4} r^2
\sin^2\theta}{F_{\sigma z}} \Bigg{[} 1 + 2 \Bigg{\{} \frac{F_{\sigma
z} H_{F1}}{2} +  \gamma_1 + \frac{H_{D0} H_{D4} r^2 \sin^2 \theta}{2
F_{\sigma z}} \Bigg{\}} \frac{\delta F_{tz}}{F_{\sigma z}}\Bigg{]}.
\end{equation}
Finally, from Noether's theorem and our ansatz above we find the angular momenta
\begin{eqnarray}
j_1 &=& \int dz d\sigma \frac{\partial {\cal L}}{\partial \dot{\phi}_1} = q_{D0} q_{F1}, \quad {\rm and} \\
\label{p-phi} j_2 &=& \int dz d\sigma \frac{\partial {\cal
L}}{\partial \dot{\phi}_2}  =  (2\pi)^2 R_A \tau_{D2} \\[2mm]
&\times& \left[r_{D4}^2 \cos^2 \theta + \frac{H_{D4} \cos^2 \theta
\dot{\phi}_2}{F_{\sigma z}} \left(F_{\sigma z}^2 H_{F1} r^2 +2
F_{\sigma z} \omega \sin^2 \theta + H_{D0} H_{D4} r^4 \sin^2
\theta\right) \right], \nonumber
\end{eqnarray}
associated with $\phi_1$ and $\phi_2$.

Now, the dynamical variables in the action (\ref{embed-action}) are
$r,\theta, \phi_2, F_{\phi z}$ and $F_{t z}$ (we have already set
$F_{t \sigma} = 0$). Recall that the conservation laws for the D0
and F1 charges provide two integrals of the motion.   Further
conservation laws follow from the three Killing vectors $
\partial_{\phi_1}, \partial_{\phi_2}$ and $\partial_t$ of
the background (\ref{t-dual-metric}), though conservation of the angular momentum
associated with $\partial_{\phi_1}$ turns out to be trivial for our
ansatz. Thus we have 4 (useful) conserved quantities and 5 dynamical
variables.

Before analyzing the resulting dynamics, it is useful to note that
the surface $\theta=\pi/2$ is fixed under translations of $\phi_2$.
Thus, any departure from $\theta = \pi/2$ would entail a breaking of
this symmetry and we may consistently restrict the dynamics to
$\theta= \pi/2$. In this case, we still have 4 useful conserved
quantities, but we have only 4 dynamical variables.  Thus, the
motion of the supertube reduces to quadratures.  We explore this
simple special case in section \ref{pi/2} below before moving on to
the general case in section \ref{gent}


\subsection{A special case: Collision in the $\theta = \pi/2$ plane}
\label{pi/2}

In this subsection we analyze the motion of the supertube when
restricted to move only in the $\theta = \pi/2$ plane.    Note that
$\phi_2$ drops out of the analysis entirely as choosing $\theta =
\pi/2$ forces the supertube to lie on the `axis'  where $\phi_2$ is
ill-defined. It is useful to define the quantity $ \mu_{F1} :=
q_{F1} \tau_{F1} /2 \pi \tau_{D2}$,  proportional to the  F1 charge.
One may then solve  (\ref{cF1-charge}) for  $\delta F_{tz}$ as a
function of $r$ and other quantities:
\begin{equation}
\label{solveFtz} \delta F_{tz} = F_{tz} -1 =  \frac{(\mu_{F1}
F_{\sigma z} - H_{D4} r^2) F_{\sigma z}^2}{H_{D4} (F_{\sigma z}^2
H_{F1} r^2 + 2 F_{\sigma z} \omega + H_{D0} H_{D4} r^4)}.
\end{equation}
When $F_{tz} = 1$, the right hand side of (\ref{solveFtz}) must
vanish. This happens precisely at the point where one satisfies
$F_{\sigma z} \mu_{F1} = H_{D4} r^2$, which is identical to the
relation between $r$ and $\mu_{F1}$ that would be imposed by
supersymmetry. Thus, $\delta F_{tz} =0$ at the minimum of the
effective potential for fixed $\mu_{F1}$.

Using (\ref{solveFtz}) and $\theta=\pi/2$, the
 energy constructed from (\ref{embed-action}) is
$ E = \tau_{D0} |q_{D0}| +  2 \pi R_A \tau_{F1}|q_{F1}| + \Delta E$,
where $\Delta E$ is the energy above the BPS limit:
\begin{eqnarray}\label{energy-pi-2}
\Delta E = (2 \pi)^2 \tau_{D2} R_A &\Bigg{[}& \frac{F_{\sigma z} r^2
(F_{\sigma z} \mu_{F1} - H_{D4} r^2)^2}{2 H_{D4} r^2 (2 F_{\sigma z}
\omega + F_{\sigma z}^2 H_{F1} r^2 + H_{D0} H_{D4} r^4)} \cr &+&
\frac{H_{D4} (2 F_{\sigma z} \omega + F_{\sigma z}^2 H_{F1} r^2 +
H_{D0} H_{D4} r^4) \dot{r}^2}{2 F_{\sigma z} r^2} \Bigg{]}. \ \ \ \ \
\
\end{eqnarray}

Setting $\dot{r}$ to zero in
(\ref{energy-pi-2}) we obtain the effective potential seen by the
supertube moving in the BMPV background:
\begin{equation}\label{potential}
V(r) = (2 \pi)^2 R_A \tau_{D2}  \frac{F_{\sigma z} r^2 (F_{\sigma z}
\mu_{F1} - H_{D4} r^2)^2}{2 H_{D4} r^2 (2 F_{\sigma z} \omega +
F_{\sigma z}^2 H_{F1} r^2 + H_{D0} H_{D4} r^4)} \ge 0.
\end{equation}
Note that the denominator of (\ref{potential}) is strictly positive
and does not vanish even at $r=0$.  Thus, $V$ vanishes at $r=0$ due
to the explicit $r^2$ in the numerator.  Since the potential is
non-negative, we see that it is always attractive near $r=0$.

Further from $r=0$ the behavior is determined by the relative sizes
of $F_{\sigma z}$, $\mu_{F1}$, and $r_{D4}^2$.  Note that
$\lim_{r\rightarrow 0} H_{D4} r^2 = r_{D4}^2$. Thus, for $\mu_{F1}
F_{\sigma z} > r_{D4}^2$ there is an additional zero at some $r >
0$. It is useful to express this relation in terms of the integer
charges by noting that
\begin{equation}
\label{factor}
\mu_{F1} F_{\sigma z} - r_{D4}^2 = \frac{g_A \alpha'^{3/2}}{R_A}
(q_{D0} q_{F1} - N_{D4}).
\end{equation}
Thus, an additional zero appears for $q_{D0}q_{F1} > N_{D4}.$ It
is precisely at this additional zero that one finds the static
supersymmetric configurations (with $\theta=\pi/2$) studied in
\cite{Bena:2004wt}.  Note that there are no supersymmetric
configurations at $r=0$, as the world-volume would be required to
be null.


\begin{figure}
\begin{minipage}[t]{2.8in}
\includegraphics[width=\textwidth]{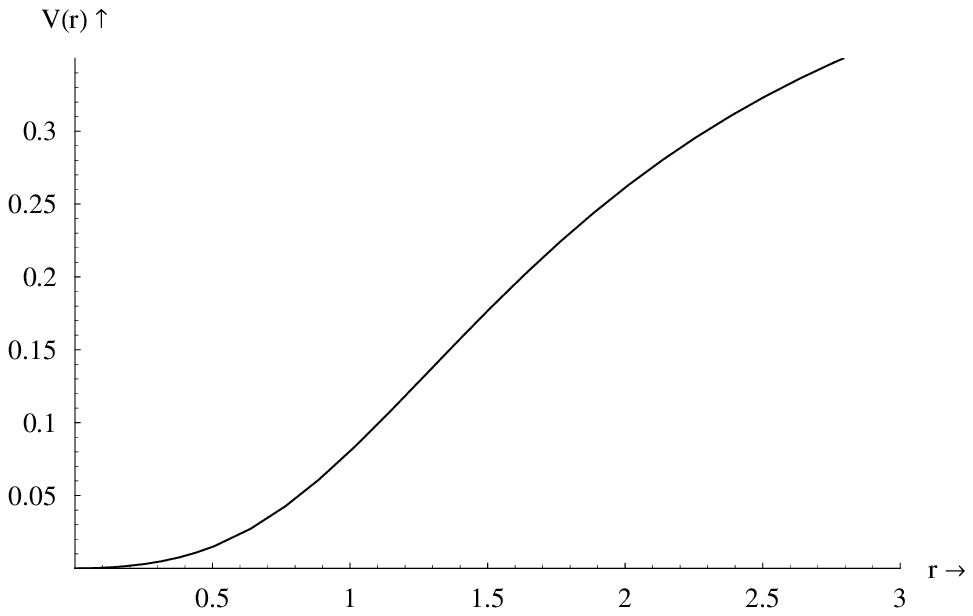} \caption[Potential I]{The potential
$\mbox{(\ref{potential})}$ with parameters $(2 \pi)^2 R_A
\tau_{D2} = r_{D0} = r_{D4} = r_{F1} = 2 \omega = 1$ and
$F_{\sigma z} = \mu_{F1} = 0.1$. Here $\mu_{F1}F_{\sigma z} -
r_{D4}^2 < 0$ and the potential is attractive for all $r$.  The
potential vanishes at $r=0$.} \label{pot-1}
\end{minipage}
\hfill
\begin{minipage}[t]{2.8in}
\includegraphics[width=\textwidth]{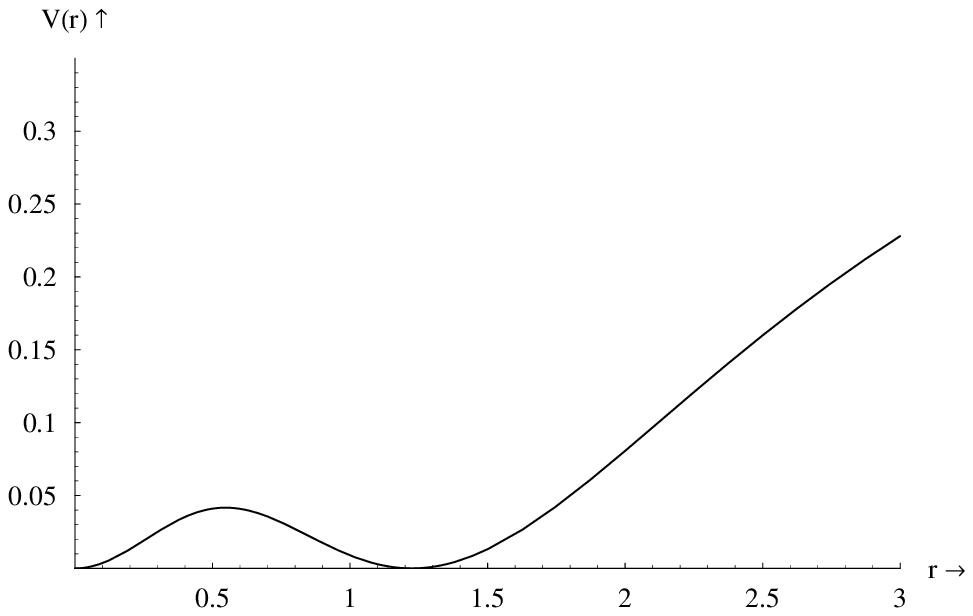} \caption[potential II]{The potential
with parameters $(2 \pi)^2 R_A \tau_{D2} = r_{D0} = r_{F1} = 2
\omega = 1, F_{\sigma z} = \mu_{F1} = 1, r_{D4} = 0.8$.  Here
$\mu_{F1}F_{\sigma z} - r_{D4}^2 > 0$ so that the potential has an
additional zero at some $r
> 0$.} \label{pot-2}
\end{minipage}
\end{figure}


We see that such solutions are separated from $r=0$ by a finite
energy barrier whose height is controlled by $q_{D0} q_{F1} -
N_{D4}$ and the various moduli.   The height of this barrier is
readily estimated by noting that the denominator of
(\ref{potential}) has its minimum at the origin, and that this
minimum is in turn greater than $2r_{D4}^4 r_{D0}^2$.  Thus we find:
\begin{equation}
\label{barrier}
Barrier \ Height \ < \tau_{D0} q_{D0} \frac{\ell^4}{(\alpha')^2} \frac{N_{D4}}{N_{D0}}.
\end{equation}
Plots
of the potential for typical cases with each sign of $q_{D0} q_{F1}
- N_{D4}$ are shown in figures \ref{pot-1} and \ref{pot-2}.

We may now ask under what circumstances a supertube will reach $r=0$
and merge with the black hole.  Let us recall from
\cite{Bena:2004wt} that, due to the fact that the supertube angular
momentum is $j = q_{D0} q_{F1}$, the sign of $q_{D0} q_{F1} -
N_{D4}$ is of interest: adding tubes with $q_{D0} q_{F1} > 4N_{D4}$
{\it can} result in an object with $(J + q_{D0} q_{F1})^2 > (N_{D0}
+ 2 q_{D0}) (N_{F1} + 2 q_{F1}) N_{D4}$ (and thus violating the BMPV
bound on angular momentum), while adding tubes with $q_{D0} q_{F1} <
4N_{D4}$ will not violate this bound\footnote{In order that the
final angular momenta will again be equal, we consider sending a
pair of otherwise identical supertubes (each with charges
$q_{D0},q_{F1}$) toward the hole, with one tube lying in the
$\phi_1$ plane and the other lying in the $\phi_2$ plane. This is
the source of factors of $2$ and $4$ above that differ from those in
\cite{Bena:2004wt}. Keeping the two angular momentum equal allows
for a more direct comparison with the BMPV bound on angular
momentum.}. In the case $q_{D0} q_{F1} - N_{D4}< 0$ (so that we also
have $q_{D0} q_{F1} - 4N_{D4}< 0$)  a supertube with $\theta =
\pi/2$ is attracted to the black hole, and we find in agreement with
\cite{Bena:2004wt} that the objects readily merge.

Suppose on the other hand that  $q_{D0} q_{F1} - N_{D4} > 0$ (as needed if we are to have
$q_{D0} q_{F1} - 4N_{D4} > 0$), and
that we begin with the supersymmetric configuration considered in
\cite{Bena:2004wt}; i.e., one located at the $r > 0$ minimum of
the potential $V$.  To cause the supertube to merge with the black
hole, we need only send it over the intervening hill. One might
imagine doing so by gently raising the supertube over the barrier
and then slowly lowering it down the other side so that it collides
with the black hole having an energy $\Delta E$ above the BPS bound that is
arbitrarily small in comparison to all other quantities.  Such a
process can lead to a final object which violates the BMPV angular
momentum bound $J^2 < N_{D0} N_{D4} N_{F1}$.  Furthermore, since the
energy above extremality is arbitrarily small compared with all
other quantities, it can also violate the similar bound which limits
the angular momentum of the non-BPS Cvetic-Youm solutions
\cite{CY1,CY2} described earlier in section \ref{prelims}.
Similarly, an object in the minimum at $r > 0$ will eventually
tunnel quantum mechanically to $r=0$ and merge with the black hole.

The purist may wish to study the case where one simply gives the
supertube a velocity toward $r=0$ which is sufficiently great to
carry it over the hill.  Again, if the black hole begins with enough
angular momentum, the final state has squared angular momentum
$(J^f)^2 = (J + q_{D0}q_{F1})^2$ larger than the product of final
charges: $N^f_{D0}N^f_{D4}N^f_{F1} = (N_{D0} + 2 q_{D0}) (N_{F1} + 2
q_{F1}) N_{D4}$.  However,  the barrier  height now sets the minimum
energy $\Delta E$ above the BPS bound for the final object.

Now, for $J_1 = J_2$ and near extremality, the
Cvetic-Youm solutions satisfy
\begin{eqnarray}
J^2_1 &<& N_{D0} N_{D4} N_{F1}   \\ &\times& \Bigl( 1 + \Delta E
\frac{2 G_5 }{\pi} \frac{r_{D4}^4 r_{F1}^4 + r_{D0}^4 r_{F1}^4
+r_{D4}^4 r_{D0}^4 + 2 r_{D0}^2 r_{D4}^2 r_{F1}^2(r_{F1}^2 +
r_{D0}^2 + r_{D4}^2)}{ r_{D0}^2 r_{D4}^2 r_{F1}^2 (r_{D4}^2 r_{F1}^2
+ r_{D0}^2 r_{F1}^2 + r_{D4}^2 r_{D0}^2)} \nonumber \cr &+&
\mathcal{O}(\Delta E)^2 \Bigr).
\end{eqnarray}
This somewhat complicated expression becomes simpler in the special
case $N_{F1} =N_{D0}=N_{D4}= N$, where it may be written
\begin{equation}
J^2_1 < N^3 \left( 1 + (moduli) \frac{ \Delta E }{N}
 + \mathcal{O}(\Delta E)^2 \right),
\end{equation}
where $moduli$ denotes a function of the moduli alone; i.e., a
factor independent of the charges $q_{D0},q_{F1}, N$.  Using
$(\ref{barrier})$ as an upper bound for $\Delta E$, it is
straightforward to arrange parameters so that the final object
violates this bound as well.   For example, for large $N$ one may
choose $q_{D0} > 2 \sqrt{N}$ and  $q_{F1} > (2 + moduli) \sqrt{N}$.
This is in contrast to the static analysis of \cite{Bena:2004wt}
which indicated that the bound on the total angular momentum would
not be violated.

In the case where we merely give the tube a finite kinetic energy,
the detailed dynamics may be read off by expanding $\Delta E$ about
$r=0$.  The leading terms are
\begin{equation}
\frac{\Delta E}{(2\pi)^2 R_A \tau_{D2}}  = \frac{(F_{\sigma z}
\mu_{F1} - r_{D4}^2)^2}{2 A} r^2 + \frac{A}{2} \frac{\dot{r}^2}{r^4} + \ higher \ order \ terms
\end{equation}
where
$$
A = \frac{r_{D4}^2 (2 F_{\sigma z} \omega  + F_{\sigma
z}^2 r_{F1}^2 + r_{D0}^2 r_{D4}^2)}{F_{\sigma z}} \ \ {\rm is  \ a \
constant.}
$$
Due to the singular kinetic term,  the supertube
reaches $r=0$ only at infinite $t$, just as one would expect for an
object falling through the horizon of a black hole.


\subsection{The general case: Motion in $\theta$}

\label{gent}


\begin{figure}
\begin{minipage}[t]{1.8in}
\includegraphics[width=\textwidth]{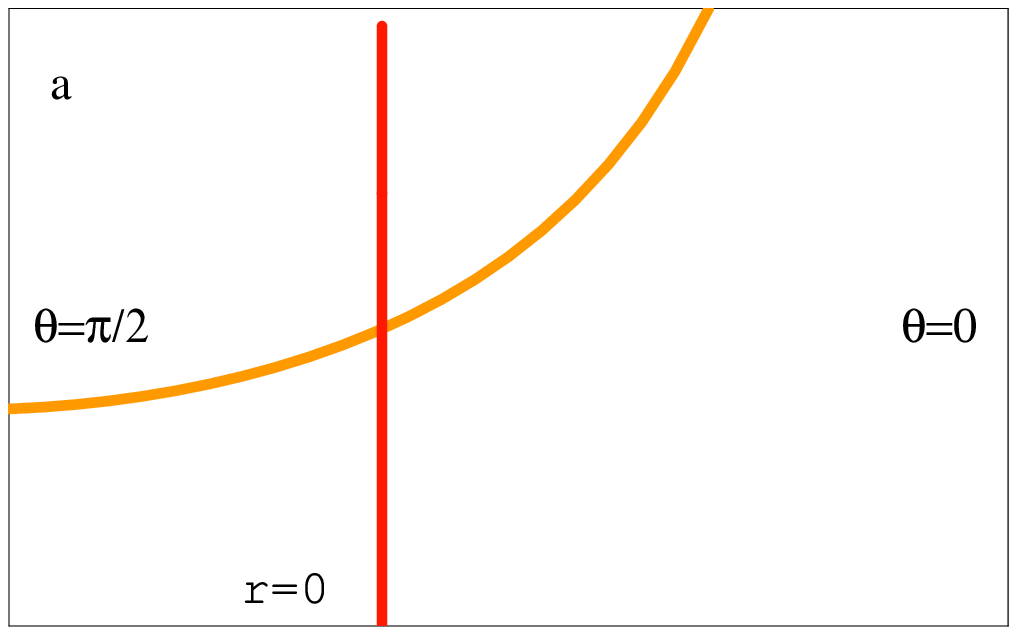}
\end{minipage}
\hfill
\begin{minipage}[t]{1.8in}
\includegraphics[width=\textwidth]{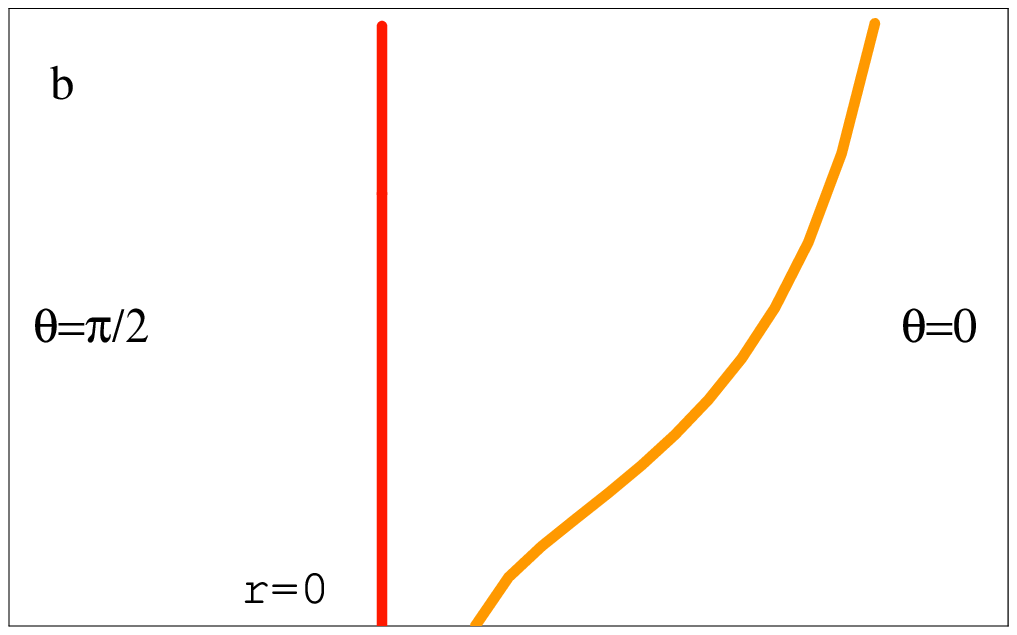}
\end{minipage}
\hfill
\begin{minipage}[t]{1.8in}
\includegraphics[width=\textwidth]{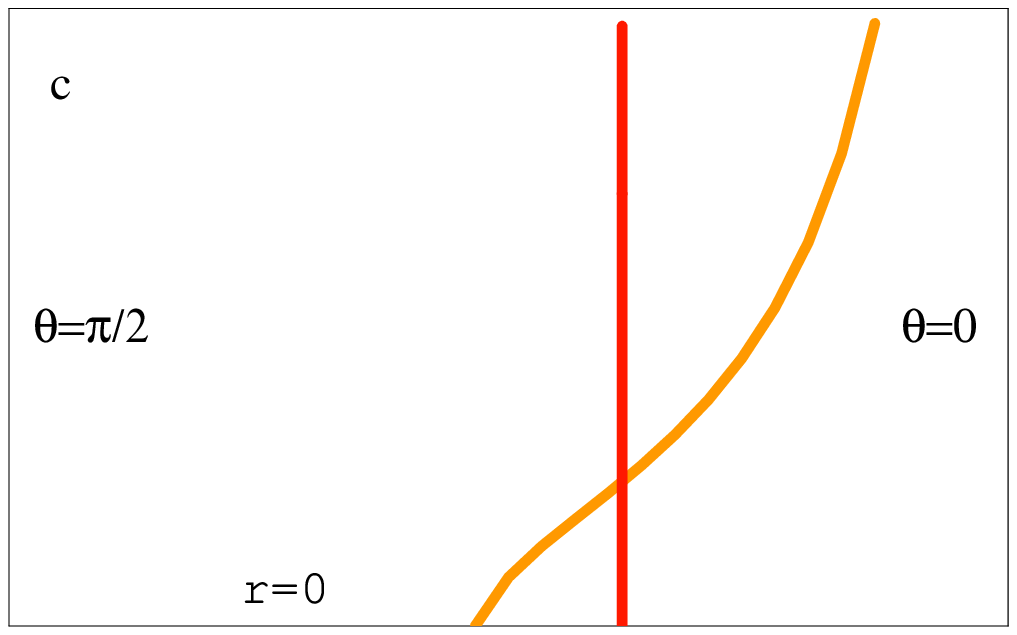}
\end{minipage}
\caption{The curves along which the two terms in (4.15) vanish are
drawn in the $(r,\theta)$ plane, with $\theta$ ranging over $[0,
\pi/2]$ from right to left on each diagram and $r$ increasing toward
the top from zero at the bottom.. The straight vertical line shows
the zeros of the second term; the location of this curve is
controlled by $j_2$. The structure of the other curve is controlled
by $q_{D0}q_{F1}-N_{D4}$.  For $q_{D0}q_{F1} > N_{D4}$, the two
curves always intersect as shown in figure a. For $q_{D0}q_{F1} <
N_{D4}$, the curves may (figure b) or may not intersect (figure c)
depending on the value of $j_2$.} \label{last-figure}
\end{figure}

The special case $\theta = \pi/2$ is useful to build intuition, but
has the disadvantage that (since motion is allowed only in $r$) the
size of the supertube is forced to change if it moves.  Since
supertubes have a preferred size this tends to create a confining
potential.  Thus we may hope that by relaxing the restriction on
$\theta$ we might find solutions in which the supertubes propagate
freely from far away to the horizon of the black hole (without our
needing to apply additional forces). Indeed, one might expect such a
result from the fact that \cite{Bena:2004wt} found BPS
configurations of fixed (small) charge arbitrarily close to the
horizon.  However, we will see that is in fact {\it not} the case
due to magnetic interactions between the black hole and the
supertube.

Proceeding along the same lines as above we find that the energy
(see (\ref{embed-action})) associated with the motion of the
supertube is $ E = \tau_{D0} |q_{D0}| + \tau_{F1}|q_{F1}| 2 \pi R_A
+ \Delta E$, where
\begin{eqnarray}\label{energy-neq-pi-2}
\Delta E &=& (2 \pi)^2 R_A \tau_{D2} \cr &\times& \Bigg{[}
\frac{H_{D4} (2 F_{\sigma z} \omega \sin^2\theta + F^2_{\sigma z}
H_{F1} r^2 + H_{D0} H_{D4} r^4 \sin^2 \theta)}{2 F_{\sigma z} r^2}
(\dot{r}^2 + r^2 \dot{\theta}^2 + r^2 \cos^2 \theta \dot{\phi_2}^2 )
\cr
 & + & \frac{F_{\sigma z} (F_{\sigma
z} \mu_{F1} - H_{D4} r^2 \sin^2 \theta)^2}{2 H_{D4} \sin^2 \theta
(F_{\sigma z}^2 H_{F1} r^2 + 2 \omega F_{\sigma z} \sin^2 \theta +
H_{D0} H_{D4} r^4 \sin^2 \theta)} \Bigg{]}.
\end{eqnarray}
Here $\dot{\phi}_2$ can be replaced with $j_2$, the conserved
angular momenta conjugate to $\phi_2$ from (\ref{p-phi}). We obtain
an effective potential by setting $\dot{r}, \dot{\theta}$ to zero in
(\ref{energy-neq-pi-2}):
\begin{eqnarray}\label{pot3D}
 V(r,\theta) &=& \frac{(2 \pi)^2  R_A \tau_{D2} F_{\sigma z} }{2H_{D4}(F_{\sigma z}^2 H_{F1} r^2 + 2 \omega F_{\sigma z} \sin^2 \theta +
H_{D0} H_{D4} r^4 \sin^2 \theta)} \cr \cr&\times& \left[
 \frac{(F_{\sigma
z} \mu_{F1} - H_{D4} r^2 \sin^2 \theta)^2}{\sin^2 \theta} +
\frac{(j_2/(2\pi)^2 R_A \tau_{D2} - r_{D4}^2 \cos^2 \theta)^2
}{\cos^2 \theta} \right].
\end{eqnarray}

\begin{figure}
\begin{center}
\includegraphics[width=7cm]{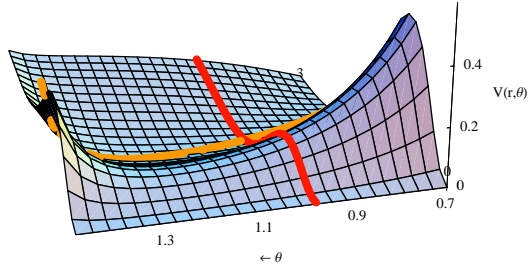} \caption{The potential as a
function of $r,\theta$ with parameters set to $(2 \pi)^2 R_A
\tau_{D2} = r_{D0} = r_{D4} = r_{F1} = 2 \omega = 1, F_{\sigma z} =
1, \mu_{F1} = 2.5, $ and $j_2 = 0.7$. The coordinate $r$ again
ranges from 0 to 3 and $\theta$ ranges from $0.7$ to $\frac{\pi}{2}
- 0.1$. The heavy lines correspond to the curves shown figure 3a.}
\label{pot2-3D}
\end{center}
\end{figure}

Note that the first term inside the square brackets will always
vanish on some curve in the $(r, \theta)$ plane, while for small
enough $j_2$ the second term will vanish at some value of $\theta$
(i.e., on another curve in the $(r,\theta)$ plane).  The qualitative
features of the dynamics are controlled by these two curves, for
which three typical configurations are sketched in figure
\ref{last-figure} below.  For $q_{D0}q_{F1}>N_{D4}$ the curves will
intersect as shown in figure \ref{last-figure}a. At the intersection
point the potential will vanish and have a local minimum. This
minimum corresponds to the general static supersymmetric supertube
considered in \cite{Bena:2004wt}. Note that the potential also
vanishes at $r=0$.  Such a potential is plotted in
Fig.~\ref{pot2-3D} for typical values of the parameters.

In contrast, for $q_{D0}q_{F1} < N_{D4}$ the curves may
(figure~\ref{last-figure}b) or may not intersect
(figure~\ref{last-figure}c), depending on the value of $j_2$.  In
the case with no intersection, the potential vanishes only at $r=0$.


\section{Discussion:  The final state} \label{discussion}

In the preceding sections we have studied the moduli-space
scattering of a two-charge supertube moving in the  type IIA
background (with F1, D0, and D4-brane charges) dual to that of a
rotating BPS D1-D5-P black hole
\cite{Tseytlin:1996as,Breckenridge:1996is}, also called a BMPV black
hole after the authors of \cite{Breckenridge:1996is}.  For
simplicity, we restricted the analysis (as in \cite{Bena:2004wt}) to
tubes preserving the $\phi_1$ translation symmetry of the background
(\ref{bmpv-field1}).  Such moduli space analyses take into account
magnetic forces, which can play important roles in the dynamics.

Although we work near the BPS limit, we found that the motion of
such tubes can be described by an effective potential $V$ depending
on two coordinates $(r, \theta)$.  This effective potential is a
result of the various conserved charges in the problem, and arises
in a manner similar to the effective potential found by Peeters and
Zamaklar \cite{PZ} in their study of a field theoretic problem
related by dualities to the scattering of two supertubes.

In our context, BPS supertube configurations arise at local minima
of this potential (with $V=0$).  Though the potential also vanishes
at the horizon $(r=0)$, magnetic forces cause these zeros to be
separated by a potential barrier.  Thus, any BPS supertube is stable
to a sufficiently small perturbation and the perturbation does not
result in merger of the tube with the black hole. However, since the
potential barrier is of finite height, the supertube may be lifted
over the barrier by additional forces (or may quantum mechanically
tunnel through) so that the tube merges with the black hole having
arbitrarily small energy above the BPS bound. This is in contrast to
the conclusion of \cite{Bena:2004wt} (based on an analysis of
exactly supersymmetric configurations) that only supertubes with
$q_{D0}q_{F1} < N_{D4}$ could merge with the black hole.  Here we
remind the reader that $q_{D0},q_{F1}$ are integer charges carried
by the supertube while $N_{D4}$ is an integer charge carried by the
black hole.

The result of such a merger is of course constrained by conservation
laws.  We considered supertubes with $j_2=0$, but $j_1\neq 0$ while
the black hole background has $J_1 = J_2$.  Since for motion on
moduli space there is no radiation
 at leading order in velocities \cite{Manton:1981mp, Ferrell:1987gf},
 the angular momenta of the object resulting from the merger are
clearly not equal.
  Thus, the final state cannot be just another
BMPV black hole. In addition, because supertubes with $q_{D0}q_{F1}
> 4 N_{D4}$ can merge with the black hole while having arbitrarily small
energy above the BPS bound,  we find in contrast to
\cite{Bena:2004wt} that, depending on both the supertube and the
original black hole,  the object resulting from the merger may also
violate the upper bound (\ref{kerr-like}) on the magnitude of the
angular momentum associated with BMPV black holes.

Now, due to the kinetic energy inherent in our scattering
analysis, the final state will not saturate the BPS bound.  Thus,
we should consider the non-extreme generalizations of BMPV (i.e.,
the Cvetic-Youm solutions \cite{CY1, CY2}  described in section
\ref{prelims}) as potential final states.  Note, however, that for
any angular momentum the departure of our supertube from the BPS
bound may be arbitrarily small in comparison with its angular
momentum (which does not vanish in the BPS limit). As a result, we
can also arrange to violate the constraints on both $J_1-J_2$ and
on $J_1 + J_2$ associated with non-extreme black holes.  Thus, a
Cvetic-Youm solution cannot by itself form the final state of our
collision.

Now, it is plausible that the Cvetic-Youm black holes are the only
black hole solutions that include  small deformations of
BMPV\footnote{Though it is also plausible that there are other
non-extreme black holes with less symmetry as was conjectured in
\cite{Harvey}. If stable, such black holes would provide additional
candidates for the final state resulting from our collision.}. Under
such an assumption, we would be forced to conclude that sending an
arbitrarily small supertube into our black hole does not result in a
small change. Such a situation might arise if our supertube triggers
an instability of the above family of black hole solutions.  While
one would expect BPS solutions like BMPV to be at least marginally
stable at linear order, the non-extreme solutions of \cite{CY1,CY2}
could well be subject to linear instabilities. Such an instability
arbitrarily close to the BPS limit is reminiscent of (and might
perhaps be connected to) the gyrational instability of the 5+1
dimensional BMPV black string suggested in \cite{gyro}.

The activation of an instability would take us out of the regime in
which the scattering is described by motion on moduli space and
could lead to a variety of effects.  For example, the activation of
an instability might lead to radiation of conserved quantities to
infinity. However, noting that the massless radiation carries no
charge, we see that the energy carried to infinity will be small
since the mass of the remaining object must still satisfy the BPS
bound.  Furthermore, radiation can carry angular momentum only if it
breaks the corresponding rotational symmetry.  Since our supertubes
preserve rotational symmetry in $\phi_1$, the radiation can change
only  the angular momentum in the $\phi_2$ plane.  For $j_1 > 0$,
such radiation might restore self-duality of the black hole's
angular momentum only by leading to an increase in the total angular
momentum.  But if the final state were to be a BMPV black hole, such
an increase of angular momentum would decrease the horizon area.  It
is therefore forbidden by the area theorem and the second law of
thermodynamics. Thus, at least for $j_1 > 0$, radiation alone does
not hold the key to the final state.

Another possibility is that the instability may lead to a
non-perturbative change in the black hole.  For example, the
instability might in principle cause the spherical ($S^3$) horizon
of the BMPV hole to become $S^1 \times S^2$, leading to a black
ring.  This is analogous to one of the scenarios discussed by
Emparan and Myers in \cite{Emparan:2003sy} with regard to their
instability of ultraspinning black holes in $ d \ge 6$ spacetime
dimensions.

Now, a final state with a black ring alone is again forbidden by the
area theorem. The point here is that, as noted in section \ref{prelims}, there is an ``entropy gap''
between the entropy of a BMPV black hole and the entropies of nearby
black rings.  Thus, for any BMPV black hole, there is an open set in
the space of conserved charges  such that every
black ring in this open set has smaller entropy than does the
original black hole.  As a result, merging a sufficiently small supertube
with the black hole cannot result in a black ring.

However, one should again allow for the effects of radiation
emitted during the supposed transition.  Let us first consider the
case where the original black hole has large angular momentum, so
that its area is nearly zero.  Then since $N_{D1} N_{D5} N_{p} -
{\cal N}_{D1}{\cal N}_5{\cal N}_{p}$ is quadratic in the dipole
charges for small dipoles whereas the coefficient of $J_1$ is
cubic, it is clear from (\ref{Sring}) that by radiating a
sufficient amount of angular momentum $J_2$ we may arrive at a
black ring of larger entropy (and with small dipole charges).
Thus, if the original black hole angular momentum is large (as in
cases leading to a violation of the BMPV angular momentum bound
(\ref{bound})), the merger may result in a black ring together
with some radiation at infinity.

On the other hand, if the angular momentum of the original black
hole is small, its entropy will be greater than the maximum entropy
(\ref{max-entropy}) of a black ring with the same conserved charges.
Thus, for this case the area theorem again forbids the final state
from being a single black ring with radiation at infinity.

Nonetheless, we note that a {\it pair} of black rings of the sort
described by Gauntlett and Gutowski \cite{Gauntlett:2004wh} could
provide the required final state. As shown in
\cite{Gauntlett:2004wh}, one can find pairs whose charges are equal
to those of a BMPV black hole but for which the total horizon area
of the pair exceeds that of the original BMPV hole by a finite
amount.  Thus, even with the small change in the charges and angular momentum associated
with the absorption of the supertube, a two black ring final state
is  entropically allowed\footnote{Cases may arise where a single
black ring together with a BMPV black hole is also allowed.}.   Thus, the merger may well
induce a fragmentation of the original black hole into multiple
black objects\footnote{The pedant may wonder what will happen if the
two black rings then collide and merge.  However, we note that any
such collision requires a breaking of the $\phi_1$ rotation
symmetry.  Thus, it cannot occur in the scenario envisioned above.
In another setting without such a symmetry, radiation could carry
away both $J_1$ and $J_2$ angular momenta, leaving behind a
slowly-rotating (or non-rotating) black hole.}.

This fragmentation scenario is also similar to one of the options
outlined by Emparan and Myers \cite{Emparan:2003sy} in discussing
their instability of ultraspinning black holes in $d \ge 6$
spacetime dimensions.  However, here we see that this is the {\it
only} available scenario based on known solutions. Thus, if our
scenario is confirmed, it will lead to a picture of black hole
interactions in $d > 4$ spacetime dimensions which is even more
similar  to that associated with particle scattering than was
previously expected. In addition to the possibility that
collisions result in the fusion of two particles into one of
greater mass, one must also allow the possibility of black hole
fission.  It would be interesting to examine this possibility
either via perturbative study of excitations about a Cvetic-Youm
black hole,  or through numerical simulations.

\subsection*{Acknowledgments}

We would like to thank Henriette Elvang, Per Kraus, Anshuman
Maharana, David Mateos, and Harvey Reall for many valuable
discussions. We also thank Iosif Bena for useful correspondence.
This work was supported in part by NSF grant PHY0354978, by funds
from the University of California, and by the Perimeter Institute
for Theoretical Physics.

\appendix

\section{The induced fields and the Dirac-Born-Infeld action}\label{appendix}

In this appendix we calculate the Dirac-Born-Infeld part of the
effective action for a supertube propagating in the type IIA dual
(\ref{t-dual-metric}) of the BMPV background
\cite{Tseytlin:1996as,Breckenridge:1996is}. We align the D2 brane
such that two of the D2-brane coordinates $ t,z $ coincide with the
spacetime coordinates,
\begin{eqnarray} t_{\rm spacetime} &=& t_{\rm supertube}
\cr z_{\rm spacetime} &=& z_{\rm supertube},
\end{eqnarray}
and restrict our analysis to circular cross sections so that  $r,
\theta, \phi_1, \phi_2$ are functions of $t$ alone.

As a first step, we calculate the fields induced on the supertube by
the embedding in our background.  We denote induced  fields by lower
case letters and background fields by upper case letters, e.g.,
\begin{eqnarray}
t_{a \ldots b} &=& \partial_a X^\mu \ldots  \partial_b X^\nu T_{\mu
\ldots \nu},
\end{eqnarray}
for a background field $T_{\mu \ldots \nu}$. In the equations below,
lower case latin indices $i,j$ run over $r,\theta,\phi_1 \mbox{ and
}\phi_2$.  We find:
\begin{eqnarray}
\label{big}
g_{00} &=& G_{00} + 2 G_{i0}\frac{\partial x^i}{\partial t} +
G_{ij}\frac{\partial x^i}{\partial t} \frac{\partial x^j}{\partial
t}= -H_{D0}^{-1/2}H_{D4}^{-1/2}H_{F1}^{-1} (1 + \gamma_t)^2+ (H_{D0}
H_{D4})^{1/2} \Delta_{tt} \cr
g_{0\sigma} &=& G_{0i}\frac{\partial x^i}{\partial \sigma} + G_{ij}
\frac{\partial x^i}{\partial t} \frac{\partial x^j}{\partial \sigma}
= - H_{D0}^{-1/2} H_{D4}^{-1/2} H_{F1}^{-1} \gamma_\sigma ( 1 +
\gamma_t) + (H_{D0} H_{D4})^{1/2} \Delta_{\sigma t} \cr
g_{0z} &=& G_{0z} + G_{iz} \frac{\partial x^i}{\partial t} = 0 \cr
g_{z \sigma} &=& G_{zi} \frac{\partial x^i}{\partial \sigma} = 0 \cr
g_{zz} &=& G_{zz} = (H_{D0} H_{D4})^{1/2} H_{F1}^{-1} \cr
g_{\sigma \sigma} &=& G_{ij} \frac{\partial x^i}{\partial \sigma}
\frac{\partial x^j}{\partial \sigma} = (H_{D0} H_{D4})^{1/2}
\Delta_{\sigma \sigma} - H_{F1}^{-1} H_{D0}^{-1/2} H_{D4}^{-1/2}
\gamma_\sigma^2
\\
b_{t\sigma} &=& B_{ti}\frac{\partial x^i}{\partial \sigma} + B_{ij}
\frac{\partial x^i}{\partial t} \frac{\partial x^j}{\partial \sigma}
= 0 \cr
b_{tz} &=& B_{tz} + B_{iz} \frac{\partial x^i}{\partial t} =
(H_{F1}^{-1} -1) + \gamma_t H_{F1}^{-1} \cr
b_{\sigma z} &=& B_{iz} \frac{\partial x^i}{\partial \sigma} =
\gamma_\sigma H_{F1}^{-1} \cr
c_{\mathit{1}t} &=& C_{\mathit{1}t} + C_{\mathit{1}i} \frac{\partial
x^i}{\partial t} = (H_{D0}^{-1} -1) + \gamma_t H_{D0}^{-1} \cr
c_{\mathit{1}\sigma} &=& C_{\mathit{1}i} \frac{\partial
x^i}{\partial \sigma} = \gamma_\sigma H_{D0}^{-1} \cr
\nonumber c_{\mathit{3}t \sigma z} &=& C_{\mathit{3}t i z}
\frac{\partial x^i}{\partial \sigma} + C_{\mathit{3}i j z}
\frac{\partial x^i}{\partial t}  \frac{\partial x^j}{\partial
\sigma} = \gamma_\sigma H_{F1}^{-1} - (H_{D4} -1) r^2 \cos^2 \theta
\left(\frac{\partial \phi_1}{\partial t} \frac{\partial
\phi_2}{\partial \sigma} - \frac{\partial \phi_1}{\partial \sigma}
\frac{\partial \phi_2}{\partial t}\right)
\end{eqnarray}
where $\gamma_\xi$ and $\Delta_{\xi \eta}$ is defined in
(\ref{gamma}) and (\ref{Delta}) respectively. Note that since we
required the velocities on the torus ($T^4$) to vanish, we have
\begin{equation}
\label{G} G_{ij} \frac{\partial x^i}{\partial \xi} \frac{\partial
x^j}{\partial \eta} = (H_{D0} H_{D4})^{1/2} \Delta_{\xi \eta} -
H_{F1}^{-1} H_{D0}^{-1/2}H_{D4}^{-1/2} \gamma_\xi \gamma_\eta.
\end{equation}

To evaluate the Dirac-Born-Infeld part of the action, we must compute  the
determinant $\sqrt{-\mbox{det}(g + b + F)}$. We wish to perform an expansion in
the velocities and the fluctuations $\delta F_{tz} :=(F_{tz} -1)$
and $F_{t \sigma}$ to second order. At zeroth order the contribution
is
\begin{equation}\label{zeroth}
H_{D0}^{-1/2}H_{D4}^{-1/2} H_{F1}^{-1} F_{\sigma z}^2,
\end{equation}
while at linear order contributing terms are
\begin{eqnarray}\label{first}
\frac{2}{H_{D4}^{1/2}  H_{D0}^{1/2} H_{F1}} \Bigg{[} F_{\sigma z}^2
\gamma_t + F_{\sigma z} H_{D0} H_{D4} \Delta_{\sigma t}  - \Bigg{\{}
F_{\sigma z}\gamma_\sigma + H_{D0} H_{D4} \Delta_{\sigma
\sigma}\Bigg{\}}\delta F_{tz} \Bigg{]},
\end{eqnarray}
and at second order we find
\begin{eqnarray}\label{second}
& - &  H_{D4}^{1/2}H_{D0}^{1/2} \Delta_{t t} \Bigg{(} F_{\sigma z}^2
+ 2 F_{\sigma z}H_{F1}^{-1} \gamma_\sigma + H_{D4} H_{D0}
H_{F1}^{-1} \Delta_{\sigma \sigma}\Bigg{)} \cr &+&
H_{D4}^{-1/2}H_{D0}^{-1/2} H_{F1}^{-1} \Bigg{[} \big{(} F_{\sigma z}
\gamma_t + H_{D0} H_{D4} \Delta_{\sigma t} \big{)}^2 +
\Bigg{(}\gamma_{\sigma}^2 - H_{D4} H_{D0} H_{F1} \Delta_{\sigma
\sigma}
  \Bigg{)}
 \delta F_{tz}^2 \Bigg{]}
\cr  & +& 2 H_{D0}^{1/2}H_{D4}^{1/2}H_{F1}^{-1} \Bigg{\{}
\Delta_{\sigma t} \Big{(}\gamma_{\sigma} + H_{F1} F_{\sigma
z}\Big{)} - \frac{\gamma_{\sigma} \gamma_{t} F_{\sigma z}}{ H_{D0}
H_{D4}} - \Delta_{\sigma \sigma} \gamma_t \Bigg{\}}\delta F_{tz} \cr
&-& H_{D0}^{1/2} H_{D4}^{1/2} H_{F1}^{-1} F_{t\sigma}^2.
\end{eqnarray}
Substituting (\ref{zeroth}--\ref{second}) into the Dirac-Born-Infeld action we find
\begin{eqnarray}\label{dbi}
S_{DBI} &=& \tau_{D2} \int dt dz d\sigma \Bigg{[}-  H_{D0}^{-1}
F_{\sigma z} - H_{D4} \Delta_{\sigma t} + \Bigg{\{} \frac{H_{D4}
\Delta_{\sigma \sigma}}{F_{\sigma z}} + \gamma_\sigma
H_{D0}^{-1}\Bigg{\}} \delta F_{tz} \cr - H_{D0}^{-1} F_{\sigma z}
\gamma_t &+& \frac{ H_{D4} H_{F1} \Sigma }{ 2 F_{\sigma
z}}\Bigg{(}\Delta_{t t} - \frac{2 \Delta_{\sigma t} \delta
F_{tz}}{F_{\sigma z} } + \frac{\Delta_{\sigma \sigma} \delta
F_{tz}^2}{F_{\sigma z}^2}\Bigg{)} + \frac{H_{D4} F_{t \sigma}^2}{2
F_{\sigma z}} \Bigg{]}.
\end{eqnarray}
Adding the Wess-Zumino terms (\ref{wess-zumino}) to $S_{DBI}$ we
obtain the full action (\ref{main}):
\begin{eqnarray}\label{full}
S &=& \tau_{D2} \int dt dz d\sigma \Bigg{[}-  F_{\sigma z} - H_{D4}
\Delta_{\sigma t} + \Bigg{\{} \frac{H_{D4} \Delta_{\sigma
\sigma}}{F_{\sigma z}}\Bigg{\}} \delta F_{tz} + \frac{H_{D4} F_{t
\sigma}^2}{2 F_{\sigma z}} - (H_{D4} -1) r^2 \cos^2 \theta \cr &
\times& \left( \frac{\partial \phi_1}{\partial t} \frac{\partial
\phi_2}{\partial \sigma} - \frac{\partial \phi_1}{\partial \sigma}
\frac{\partial \phi_2}{\partial t}\right) + \frac{ H_{D4} H_{F1}
\Sigma }{ 2 F_{\sigma z}}\Bigg{(}\Delta_{t t} - \frac{2
\Delta_{\sigma t} \delta F_{tz}}{F_{\sigma z} } +
\frac{\Delta_{\sigma \sigma} \delta F_{tz}^2}{F_{\sigma
z}^2}\Bigg{)}
 \Bigg{]}.
\end{eqnarray}


\end{document}